\documentclass[aip,apl,twocolumn,amsmath,amssymb,reprint]{revtex4}

\usepackage{graphicx}
\usepackage{dcolumn}
\usepackage{bm}

\begin{document}

\title{First-principles investigation of hydrogen adatoms on uniaxially strained graphene}
\author{Daniel Cole}
\author{Li Yang}
\affiliation{Department of Physics, Washington University, St. Louis, MO 63130, USA}

\date{\today}

\begin{abstract}
We have performed first-principles studies on adsorption patterns
of hydrogen adatoms on uniaxially strained graphene. Our
simulation reveals that the adsorption energy of adatoms are
sensitive to the strain. Hydrogen adatoms on zigzag strained
graphene tend to form a chain-like adsorption patterns
perpendicular to the strain direction, but those under armchair
strain cannot form any long-range chain pattern. We explain our
results in terms of a tight-binding model and the electronic
structure of strained graphene. These anisotropic adsorption
behaviors under uniaxial strain suggest methods for obtaining
regular adsorption patterns and tailoring the electronic structure
of graphene.
\end{abstract}

\maketitle

Graphene, a single layer of graphite, has ignited tremendous
research interest because of its unique electronic structure and
associated exciting electrical, thermal, and optical properties
\cite{Nov2004Sci, Geim2009Sci, rev-1, rev-2}. However, to realize
devices based on graphene we have to develop methods of tailoring
its electronic structure. One promising approach is to adsorb
chemical groups and adatoms onto graphene to modulate its
electronic and optical properties \cite{adsorb-1, adsorb-2,
Nov2005, Xia2010, Chen2007, Wang2009}, giving rise to the
large-scale production of functionalized graphene \cite{Yan2009,
McAllister2007, Boukhvalov2009}.  For example, a finite and
tunable band gap can be obtained by the adsorption of various
chemical groups, making it possible to use graphene for bipolar
devices \cite{Chernozatonskii2007, Duplock2004, adsorb-2}.
However, a known drawback of this approach is that these
adsorption patterns are usually irregular, resulting in unwanted
scattering, severely decreasing the mobility of free carriers.
Therefore, understanding the mechanism behind these adsorption
patterns and how to control them are crucial for developing
broader applications of graphene.

On the other hand, graphene exhibits excellent mechanical
properties; recent experiments have shown that well-prepared
graphene can sustain uniaxial strain up to 15\%, making graphene
one of the most stretchable crystalline structures known
\cite{strain-1, strain-2}. We can capitalize on the impressive
mechanical characteristics of graphene to develop ways of
modulating the electronic structure, motivating us to study the
adsorption of chemical groups or adatoms on stretched graphene.
Because of the broken symmetry, we expect that the adsorption
preference of adatoms on stretched graphene will be significantly
modified, \emph{i.e.}, the formation energy of adsorption patterns
may be strongly anisotropic. This would give hope of controlling
the adsorption process to form regular patterns.

In this Letter, we focus on the strain effects on hydrogen adatoms
because recent experiments have shown that hydrogen atoms can be
efficiently and precisely adsorbed onto graphene \cite{adsorb-1,
Casolo2009}. First-principles simulations are carried out to study
the adsorption energy of hydrogen adatoms on graphene with +5\%
stretch. We examine two basic uniaxial strain directions, armchair
and zigzag. The energetically favored adsorption patterns of
hydrogen adatoms under zigzag strain tend to form a chain-like
configuration that is always perpendicular to the strain
direction, but those under armchair strain cannot hold the regular
chain-like pattern for adatoms more than five. The regular
chain-like pattern obtained by zigzag strain can be well explained
by the tight-binding model and the randomness of the adsorption
pattern under armchair strain is attributed to the resonance
between the atomistic modulation and electrons around the Dirac
cone.

\begin{figure}
\centering
\includegraphics[scale=0.23]{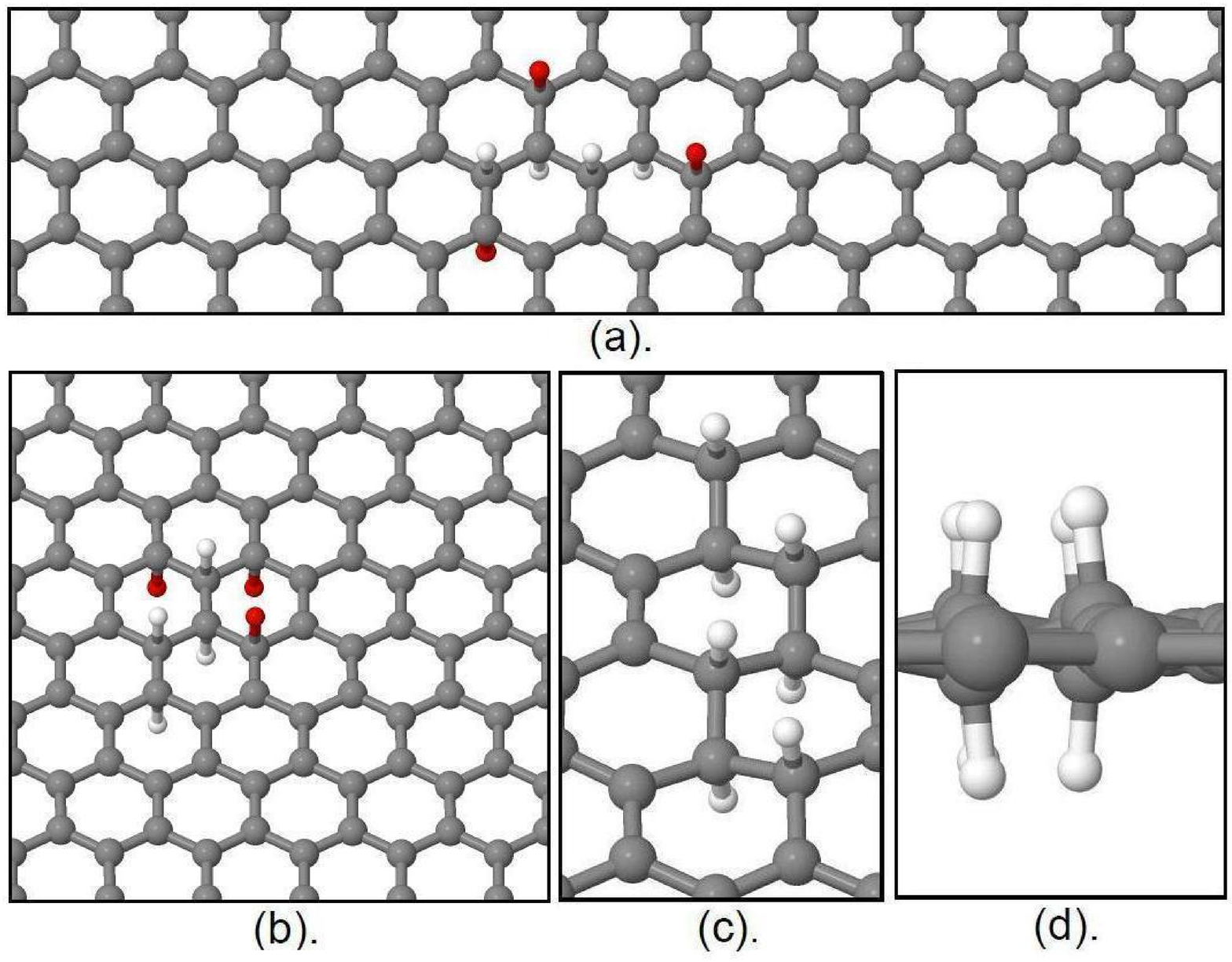}
\caption{(a) A schematic depiction of the supercell for the
armchair strained graphene. (b) That for the zigzag strained
graphene. The red atoms denote possible adsorbing sites whose
adsorption energy will be evaluated. (c) and (d) show two
perspectives on the final, seven adatom configuration with zigzag
strain. Note the structural distortion in (c) and the inward
leaning of the adatoms in (d).} \label{Figure1}
\end{figure}

We perform our simulations by employing density functional theory
(DFT) within the local density approximation (LDA) through the
Quantum Espresso simulation package \cite{Giannozzi2009} with
norm-conserving pseudopotentials \cite{Troullier1991, Perdew1981}.
To avoid artificial interactions between neighboring graphene
layers, we set the inter-layer distance to be 1.2 nm. The force
and stress are fully relaxed according to DFT/LDA. In order to
mimic isolated patterns of adsorbed hydrogen atoms, we choose
supercells of sufficient size. Here we use rectangular supercells
composed of 96 carbon atoms, but the geometry of the supercell is
different for each strain condition. As shown in Fig.
\ref{Figure1} (a), the supercell for armchair strain has 24 carbon
atoms along its zigzag axis and 4 carbon atoms along its armchair
axis. The supercell for zigzag strain shown in Fig. \ref{Figure1}
(b) has 12 carbon atoms along its zigzag axis and 8 carbon atoms
along its armchair direction. These choices of supercells are
based on our later simulation which reveals that adsorption
patterns usually prefer to be perpendicular to the strain
direction. As a result, the above supercell choices can maximize
the distance between adsorption patterns and minimize the
interaction between neighboring ones. A plain-wave basis is used
with an energy cutoff of 816 eV. A 2x2x1 k-point sampling grid is
employed for the energy convergence.

Our simulation procedure to find the energetically preferred
adsorption pattern is as follows: We affix the first hydrogen atom
above a host carbon on a graphene sheet and fully relax the
structure. Then we determine the adsorption energy of a second
adatom at each of the nearest and second nearest neighbor
positions by comparing the total energy obtained from DFT/LDA.
Actually only the nearest neighboring positions need to be
considered because our simulations and previous results
\cite{Casolo2009} show that these nearest neighbor positions are
always more energetically preferred. Moreover, the nearest
neighbor adatoms prefer to alternate above and below the graphene
layer because this conforms to the sp$^3$ hybridizing
configuration of the host carbon atom and minimizes structural
distortion. After finding the energetically preferred position for
the second adatom, we can continue this search procedure to find
those positions for the third and fourth adatoms and so on. We
illustrate two examples of this process in Figs. 1 (a) and (b),
where four hydrogen atoms shown in white are fixed adsorption
adatoms and those in red are the possible sites we consider for
the fifth adatom. Because of the symmetry, other nearest neighbor
sites are symmetrically equivalent to the three shown in Figs. 1
(a) and (b)

\begin{figure*}
\centering
\includegraphics[scale=0.40]{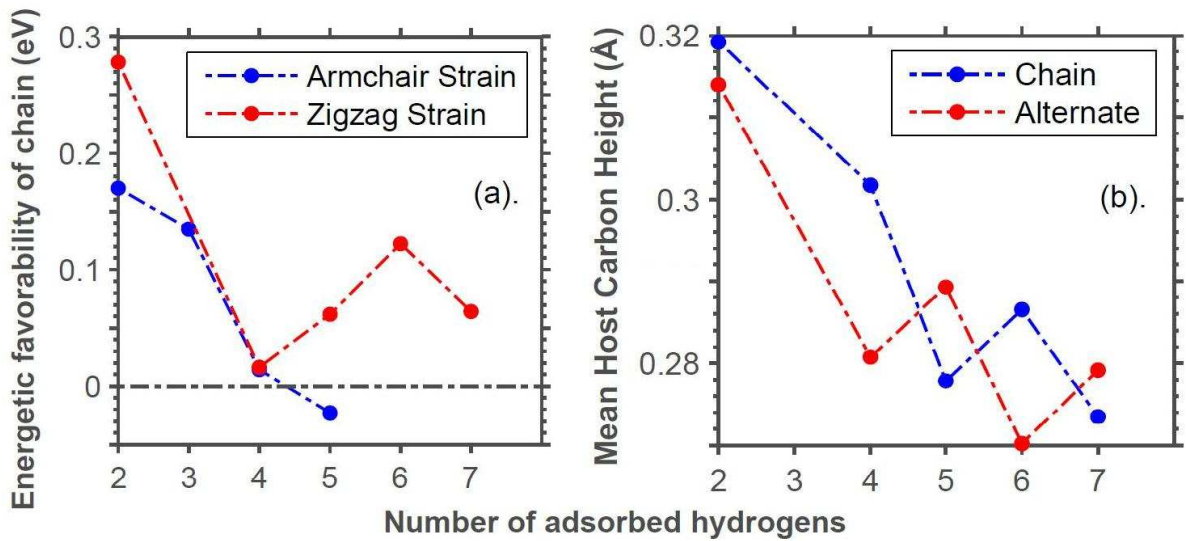}
\caption{(a) The plot of the relative energetic favorability of
the chain configuration relative to lowest energy alternative
configuration. Note that at five adatoms, the armchair
strained system no longer prefers growth perpendicular to the
strain.  (b) The plot of the mean of the absolute vertical
deviation of host carbon atoms from the graphene plane for zigzag
strain. The chain configuration is always preferred; this plot
illustrates the alternating influences of dimerization and
minimization of structural distortion, as discussed above.}
\label{Plots}
\end{figure*}

We have investigated the adsorption energy for two types of
strain, zigzag and armchair, respectively. The adsorption energies
of patterns under these two strain conditions exhibit
qualitatively different behaviors as shown in Fig. \ref{Plots}
(a), where we plot the energy difference between the chain
configuration, which is defined by the regular pattern
perpendicular to the strain, and the alternative configuration
with the lowest energy.

In the zigzag strain case, we check the adsorption energy up to
seven hydrogen adatoms. The energetic favorability of the chain
configuration is always positive, meaning the chain configuration
along the armchair direction is the energetically preferred
adsorption pattern, depicted in Figs. \ref{Figure1} (c) and (d).
For most cases, the adatom adsorbed along the armchair direction
is preferred by around one hundred meV. This values of the energy
gain could be tuned by choosing other chemical adsorption groups,
which is important for the self assembly of these patterns. This
general preference to grow a pattern perpendicular to the strain
can be understood in terms of the change in the orbital
hybridization upon adsorption. After hydrogen atoms are adsorbed
onto graphene, the hybridization of host carbon atoms changes from
sp$^2$ to sp$^3$, and the bond length between the nearest
neighboring carbon atoms is extended. As a result, the
perpendicular adsorption line is the best way to release the
applied strain and lower the total energy.

Since the adsorption of hydrogen atoms on graphene involves a
change in hybridization of host carbon atoms from sp$^2$ to
sp$^3$, it is of interest to study how the local chemical bonds
change with the adsorption process. We plot the mean of the
absolute deviation of host carbon atoms from the graphene plane
for both the chain configuration and the second lowest energy
configuration for zigzag strain in Fig. \ref{Plots} (b). An
interesting oscillating behavior is observed for the average
height of host carbon atoms. We can explain this behavior in terms
of a tight-binding model (see Refs. 22, 23 for details of this
model) as the following.

\begin{figure}
\centering
\includegraphics[scale=.22]{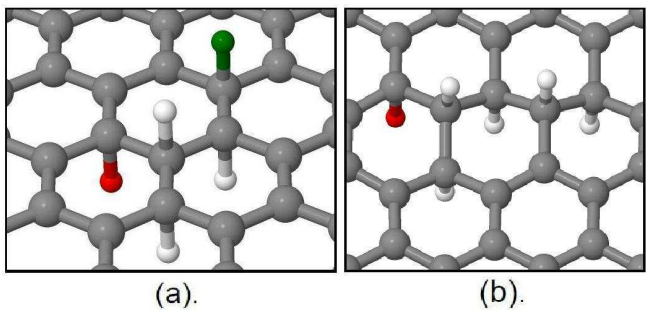}
\caption{(a) Two adatom placements whose energies differ by only
16 meV due to the competing influences of dimerization and
minimization of structural distortion.  The green adatom has lower
energy due to the dimerization effect.  (b) The preferred
configuration for 5 adatoms on armchair strained graphene.  The
red adatom highlights the expected location of the fifth adatom
without supercell resonance effects.} \label{Figure3}
\end{figure}

When graphene is strained along the zigzag axis the C-C bonds
along the zigzag bonds increase in length, while that along the
armchair axis decreases in length after the full structural
relaxation. Correspondingly, the tight-binding hopping parameters
along the zigzag direction decrease, while that along the armchair
direction increases. This leads to a dimerization of C-C bonds
along the armchair direction. When the number of hydrogen atoms
present on the graphene sheet is odd there is always a dimer with
a single orphaned electron, and it is energetically preferable for
the subsequent hydrogen to adsorb to the other carbon to form the
dimer pair, substantially increasing the energy gain for
adsorption patterns along the armchair direction. When an even
number of hydrogen atoms are present, however, there is no such
effect, and the placement of the next adatom is instead determined
by reduction of structural distortion, which is a small energy
gain. This explains the alternating pattern of the plot in Fig.
\ref{Plots} (b). This interpretation also sheds light on the
notably small energetic favorability of the chain when there are
four adsorbed hydrogens as shown in Fig. \ref{Plots} (a). In this
case because the chain is short, the net structural distortion is
very sensitive to the placement of the fourth adatom, and this
factor competes well with the dimerization effect. The two
competing systems are shown in Fig. \ref{Figure3}(a).

Meanwhile, we observe a strikingly different behavior in the case
of armchair strain; a chain of adsorbed hydrogen adatoms begins to
form along the zigzag direction. However, once the length of the
chain reaches five atoms, the zigzag direction is no longer the
energetically preferred configuration and the system depicted in
Fig. \ref{Figure3}(b) is preferred. Even more interesting is that
the critical length of the chain is dependent upon the size of the
supercell. In a supercell which is nine carbon atoms wide in the
zigzag direction, the chain can only be held for up to three
adatoms, while for a ten carbon wide supercell the chain can be
held for up to six adatoms.

This supercell effect can be attributed to the interaction between
adsorption patterns with the electrons around the Dirac cones
where the Fermi level is located. In graphene, those active
electrons at the Fermi level have a characteristic wavelength
determined by the position of the Dirac cones in the first
Brillouin zone. As we vary the size of the supercell and the
adsorption patterns, they can resonant with the Dirac electrons if
the spacial frequency of the adsorption pattern has a component
close to the wavelength of the Dirac cones. This unusual formation
energy change only occurs for the zigzag-directed pattern because
the Dirac cones in our rectangular supercell lie along the zigzag
axis \cite{Pereira2009}. Although this supercell effect is due to
the limitations of periodic boundary conditions, it is of
importance in the realistic dynamic self assembly of these
patterns because their periodicity will be influenced by the same
mechanism.

In summary, through first-principles calculations we have found
that it is possible to direct adsorption patterns of hydrogen on
graphene with the application of uniaxial zigzag strain. Although
the energy gain is not extremely significant, it could be improved
by choosing other chemical groups with a smaller binding energy.
The results in the armchair case are not as positive due
apparently to the limitations of periodic boundary conditions and
associated enhanced perturbation from electrons around Dirac
cones.

Support from the International Center for Advanced Renewable
Energy and Sustainability (I-CARES) and the Delos Fellowship at
Washington University is gratefully acknowledged. We acknowledge
computational resources support by the Lonestar of Teragrid at the
Texas Advanced Computing Center (TACC) and the National Energy
Research Scientific Computing Center (NERSC) supported by the U.S.
Department of Energy.


\end{document}